\documentclass[useAMS,usenatbib,usegraphicx]{mn2e}

\newcommand{\kms}{km s$^{-1}$}
\newcommand{\cmN}{cm$^{-2}$}

\newcommand{\lam}{$\lambda$}
\newcommand{\civ}{\mbox{C\,{\sc iv}}}

\newcommand{\siiv}{\mbox{Si\,{\sc iv}}}
\newcommand{\siiii}{\mbox{Si\,{\sc iii}}}
\newcommand{\siv}{\mbox{S\,{\sc iv}}}

\newcommand{\ovi}{\mbox{O\,{\sc vi}}}
\newcommand{\nv}{\mbox{N\,{\sc v}}}

\newcommand{\pv}{\mbox{P\,{\sc v}}}
\newcommand{\feii}{\mbox{Fe\,{\sc ii}}}

\title[A variable \pv\ BAL and quasar outflow energetics]
{A variable \pv\ broad absorption line and quasar outflow energetics}
\author[Capellupo, Hamann \& Barlow]{D. M. Capellupo$^{1,2}$
\thanks{E-mail:danielc@wise.tau.ac.il (DMC)}, F. Hamann$^{1}$, T. A. Barlow$^3$\\
$^{1}$Department of Astronomy, University of Florida, Gainesville, FL 32611-2055\\
$^{2}$School of Physics and Astronomy, Tel Aviv University, Tel Aviv 69978, Israel\\
$^{3}$Infrared Processing and Analysis Center, California Institute of Technology, Pasadena, CA 91125}

\begin{document}


\pagerange{\pageref{firstpage}--\pageref{lastpage}} \pubyear{2002}

\maketitle

\label{firstpage}

\begin{abstract}
Broad absorption lines (BALs) in quasar spectra identify high velocity outflows that might exist in all quasars and could play a major role in feedback to galaxy evolution. The viability of BAL outflows as a feedback mechanism depends on their kinetic energies, as derived from the outflow velocities, column densities, and distances from the central quasar. We estimate these quantities for the quasar, Q1413+1143 (redshift $z_e = 2.56$), aided by the first detection of \pv\ \lam1118, 1128 BAL variability in a quasar. In particular, \pv\ absorption at velocities where the \civ\ trough does not reach zero intensity implies that the \civ\ BAL is saturated and the absorber only partially covers the background continuum source (with characteristic size $<$0.01 pc). With the assumption of solar abundances, we estimate that the total column density in the BAL outflow is $\log N_H \ga 22.3$ \cmN. Variability in the \pv\ and {\it saturated} \civ\ BALs strongly disfavors changes in the ionization as the cause of the BAL variability, but supports models with high-column density BAL clouds moving across our lines of sight. The observed variability time of 1.6 yr in the quasar rest frame indicates crossing speeds $>$750 \kms\ and a radial distance from the central black hole of $\la$3.5 pc, if the crossing speeds are Keplerian.
The total outflow mass is $\sim$4100 $M_{\sun}$, the kinetic energy $\sim$$4 \times 10^{54}$ erg, and the ratio of the outflow kinetic energy luminosity to the quasar bolometric luminosity is $\sim$0.02 (at the minimum column density and maximum distance), which might be sufficient for important feedback to the quasar's host galaxy.

\end{abstract}


\section{Introduction}

Broad absorption lines (BALs) in quasar spectra identify high velocity outflows that originate from the quasar accretion disk and could play a major role in feedback to galaxy evolution. In order to determine the viability of BAL outflows as a feedback mechanism, we need estimates of their mass outflow rates and kinetic energy yields. These quantities depend on the outflow speeds, which are easy to measure, plus the column densities and distances of the flows from the central SMBH, which are much more difficult to derive. Distances can be inferred in some cases from BAL variability \citep{Misawa07,Moe09,Capellupo11,Capellupo13,Hall11,RodriguezH11,RodriguezH13} or from excited state lines with photoionization modeling \citep{Moe09,Dunn10,Borguet13}. Column density estimates are hampered by saturation. In an unknown number of cases, the apparent optical depths of BAL troughs give just lower limits on the true optical depths and column densities because the absorbing gas only partially covers the background light source (e.g. \citealt{Hamann98,Arav99a,Gabel03}).

One way to overcome the problem of saturation in BALs is to search for absorption in low-abundance ions, such as \pv\ (P/C $\sim$ 0.001 in the Sun; \citealt{Asplund09}). \pv\ is a good choice because it has a resonance doublet at accessible wavelengths, 1118 and 1128 \AA, and its ionization is similar to much more abundant and commonly measured ions such as \civ\ \lam\lam1548,1551. \pv\ absorption should be present if the column densities in the outflows are large enough. Originally, detections of \pv\ absorption were interpreted as indications of a vast over-abundance of phosphorus (e.g. \citealt{Turnshek88}; \citealt{Junkkarinen97}; \citealt{Hamann98}). The main problem is a poor understanding of the true BAL optical depths due to partial covering of the background continuum source (\citealt{Hamann98,Arav99a,Hamann02,Gabel03}). However, if the relative abundances are roughly solar, the presence of \pv\ absorption implies that \civ\ and other common BALs are extremely optically thick and the total outflow column densities are much larger than previously supposed. For example, assuming solar abundances and a standard ionizing spectrum in photoionization models, \citet{Hamann98} showed that the true \civ\ optical depths in one BAL quasar are at least $\sim$800 times greater than \pv\ in idealized BAL clouds that are optically thin throughout the Lyman continuum. In other situations with total column densities up to $N_H \sim 4\times 10^{23}$ cm$^{-2}$, the ratio of \civ\ to \pv\ optical depths might be as low as $\sim$100. \citet{Leighly09} and \citet{Leighly11} present similar results across a wide range of physical conditions. Using many more observational constraints on the outflow conditions in a particular quasar, \citet{Borguet12} estimated that the \civ /\pv\ optical depth ratio should be $\sim$1200. Altogether these results show that even a weak \pv\ detection indicates a very saturated \civ\ BAL and large total column densities.

As part of our survey to study BAL variability \citep{Capellupo11,Capellupo12,Capellupo13}, we discovered variability in \pv\ and other BALs in the quasar Q1413+1143. The \pv\ absorption was detected previously by \citet{Monier98}. However, this is the first report of \pv\ BAL {\it variability}, in this or any quasar, to our knowledge. It is significant because it confirms that the measured feature is indeed \pv\ in the quasar's BAL outflow rather than a coincidental broad blend of unrelated, intervening Ly$\alpha$ absorption lines in the Ly$\alpha$ forest. It is also significant because it favors models for the variability that involve clouds crossing the line of sight at high speeds \citep{Hamann08,Hall11,Capellupo12,Capellupo13,Wildy14}. We analyze new and existing spectra of Q1413+1143 to derive new constraints on the location and column densities of the BAL outflow.

Q1413+1143, also known as the Cloverleaf, is a well-studied lensed quasar at $z_e = 2.563$ \citep{Adelman08}, with four components of similar brightness. The faintest component is within 0.5 mag of the brightest component \citep{Turnshek97}. The components have a small separation; all are within 0''.6 of the image center \citep{Magain88}. The four components are also similar spectroscopically. Although the BEL equivalent widths are smaller, and there are some small differences in the \civ\ absorption profile, in the faintest component \citep{Angonin90}, \citet{Hutsemekers10} attributes the spectral differences in this faint component to a long-term microlensing effect, which magnifies the continuum source in this component.

Section \ref{data} describes the data and properties of Q1413+1143. Section \ref{anal} discusses the variability in the \pv\ BAL and describes the constraints that the \pv\ detection places on the column density of the outflow in this quasar. Section \ref{discuss} discusses our estimates of the energetics of this flow and their implications.

\section{Data}
\label{data}

\begin{table}
    \caption{Data Summary.}
    \begin{tabular}{clccc}
\hline
Year & Telescope & Resolution & $\Delta\lambda$ & $\Delta$t  \\
  &  & (\kms)  & (\AA) & (yr) \\
\hline
    1989.26	& Lick 3-m	  & 530  & 3793--6382	 & ...		\\
    1994.98	& HST FOS	    & 250  & 3235--4781  & 1.6 yr \\
    2000.29 & HST STIS    & 560  & 2900--5709  & 3.1 yr \\
    2006.31	& SDSS 2.5-m	& 150  & 3803--9221	 & 4.8 yr	\\
    2010.11	& KPNO 2.1-m	& 280  & 3482--6232	 & 5.9 yr	\\
\hline
    \end{tabular}
    \label{tab}
\end{table}

During the BAL monitoring campaign described in \citet{Capellupo11,Capellupo12,Capellupo13}, we monitored 24 BAL quasars, including Q1413+1143, and we searched for variability in the \civ\ and \siiv\ absorption lines. In Q1413+1143, we also detected a variable \pv\ absorption line. The monitoring campaign includes 7 epochs of ground-based data for this quasar, but only 3 have wavelength coverage that extends blue enough to include the \pv\ BAL. The first of these 3 spectra is originally from \citet{Barlow93}, observed in April 1989, and the second spectrum is from the Sloan Digital Sky Survey \citep{Adelman08} and was taken in April 2006. The most recent spectrum was taken in February 2011 at the KPNO 2.1m. As mentioned earlier, Q1413+1143 is lensed. However, given the small separation between the four components, these spectra include light from all four components.

We further supplement our dataset for Q1413+1143 with spectra from the HST archive. Q1413+1143 was observed by both FOS and STIS. The FOS data has individual observations of each of the four lens components. We use the sum of the four components to compare to the observations above that include integrated light from all the components. We note that the D component of the lens had a weaker Ly$\alpha$ \lam1216 emission line than the other components (see also, \citealt{Angonin90}), but the absorption in \pv\ is identical in all the components. For the STIS data, there are separate observations of just the A and B components of the lens. The two spectra are nearly identical, so we combined them to improve the signal-to-noise.

Table \ref{tab} summarizes all of the data covering \pv\ that we discuss in this paper. The first column gives the fractional years of each observation. The next three columns list the observatories, the spectral resolution of the data, and the wavelength coverage. The final column lists the rest-frame time-scale, $\Delta$$t$, in comparison to the first observation, from 1989.26. For more details on these data, see \citet{Capellupo11} for the Lick and SDSS spectra, \citet{Monier98} for the HST FOS spectra, \citet{Monier09} for the HST STIS spectra, and \citet{Capellupo13} for the KPNO spectrum.

For some of our discussion of the BAL outflow properties below, we estimate a bolometric luminosity based on $\lambda L_{\lambda}$(1500\AA), measured from the absolute flux-calibrated Lick spectrum, in a cosmology with $H_o = 71$ \kms\ Mpc, $\Omega_M = 0.3$, and $\Omega_{\Lambda}=0.7$. Using a standard bolometric correction factor, $L\approx 4.4\lambda L_{\lambda}$(1500\AA) \citep{Hamann11}, and dividing by a lens magnification factor of 11 \citep{Venturini03}, gives a bolometric luminosity of $L = 2.4 \times 10^{46}$ ergs s$^{-1}$ for Q1413+1143.
We approximate the black hole mass by assuming $L = 1/3L_{edd}$, which gives $M_{\mathrm{BH}} \approx 4.8 \times 10^{8}$ M$_{\sun}$.

\section{Analysis and Results}
\label{anal}

\subsection{Variability in a \pv\ absorption line}
\label{var}

\begin{figure*}
  \centering
  \includegraphics[width=135mm]{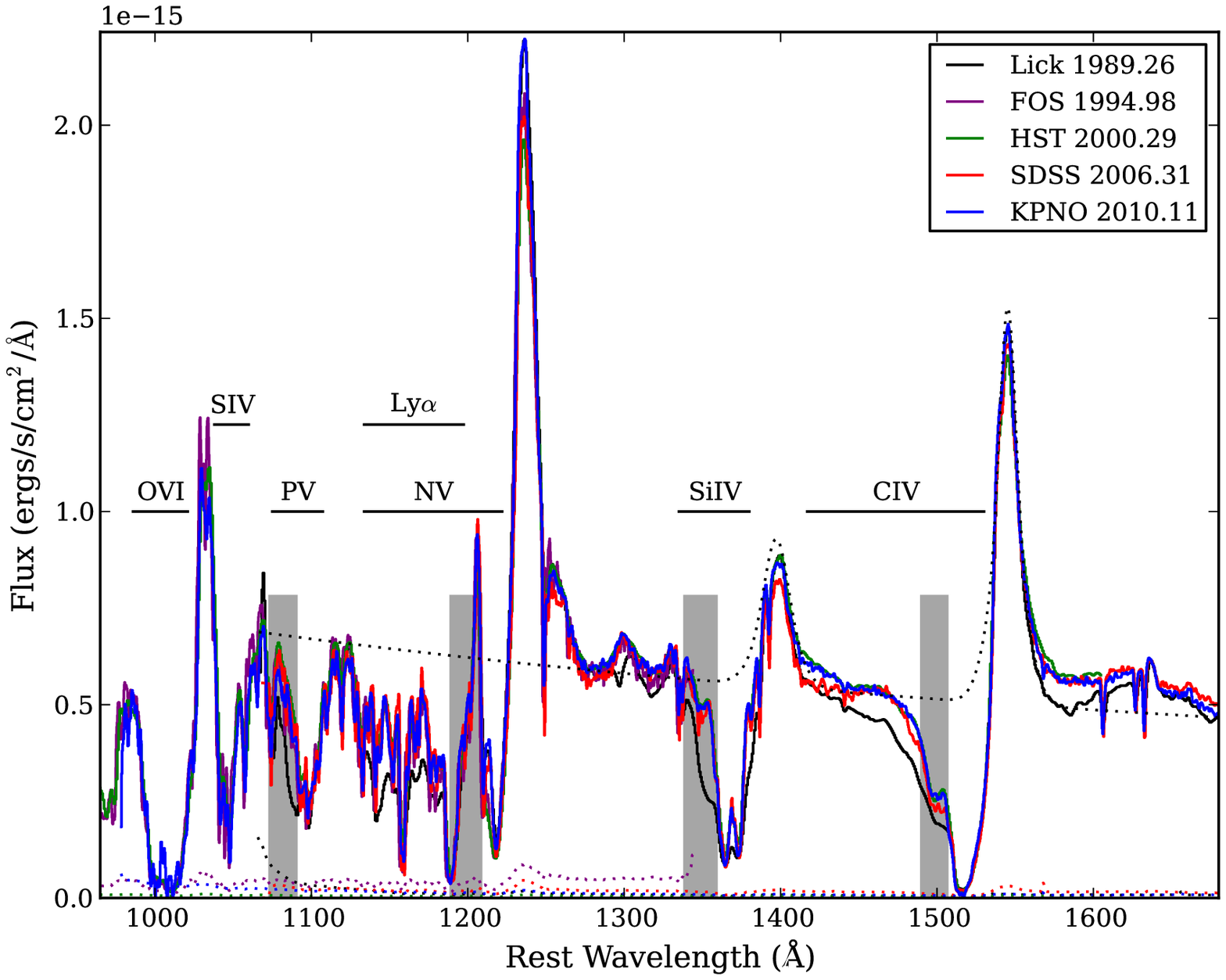}
  \caption{Spectrum of Q1413+1143, showing all of the epochs outlined in Section 2 and Table
    \ref{tab}. The middle shaded region marks the interval of variability in \siiv\ in the blue side of the trough, and, in order of increasing wavelength, the other three shaded regions show the corresponding velocity intervals in \pv, \nv, and \civ. The black dotted curve is the pseudo-continuum fit, and formal 1$\sigma$ errors are plotted across the bottom.}
  \label{spectra}
\end{figure*}

\begin{figure}
  \centering
  \includegraphics[width=75mm]{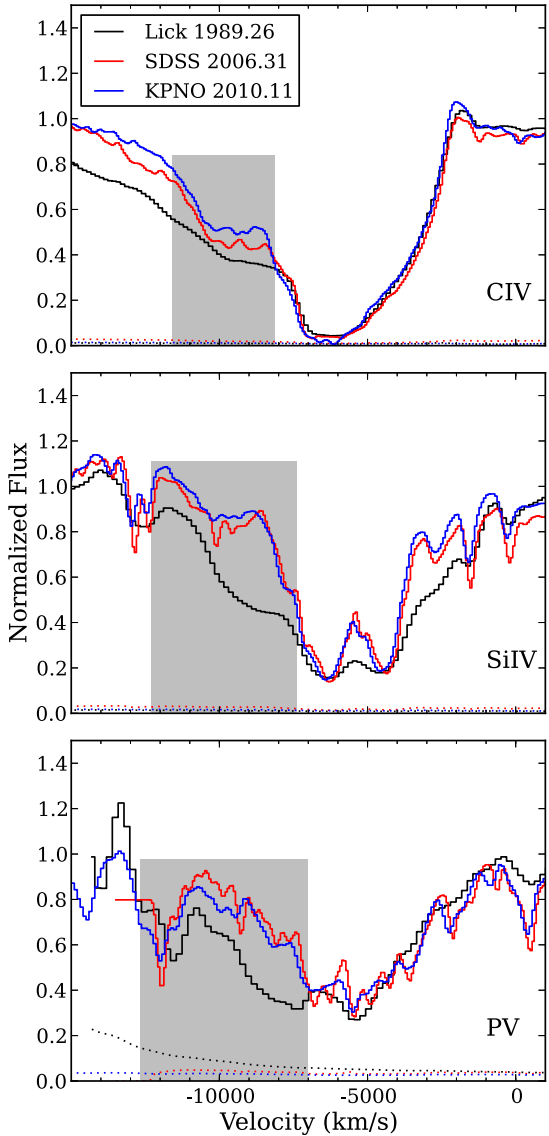}
  \caption{The BAL line profiles for, from top to bottom, \civ, \siiv, and \pv. The shaded regions here mark the same intervals as in Fig. \ref{spectra}.}
  \label{stack}
\end{figure}

In the variability study described in \citet{Capellupo11,Capellupo12}, we find that Q1413+1143 varied in both \civ\ and \siiv. We also detect the \pv\ absorption line with variability at velocities corresponding to the variability in both \civ\ and \siiv, which we discuss here for the first time. Out of the 7 epochs of data we have for this object in our BAL monitoring programme, 3 of the spectra extend blue enough to show this \pv\ absorption (Table \ref{tab}).

We adopt the Lick spectrum here as the fiducial spectrum, for which we fit a pseudo-continuum in \citet{Capellupo11}. We use the same pseudo-continuum fit in this paper. This pseudo-continuum includes a power-law fit to the continuum in regions free of emission and absorption, and for Q1413+1143, we used the spectral regions at 1270$-$1290 and 1680$-$1700 \AA. We also fit multiple gaussians to the \civ\ and \siiv\ broad-emission lines (BELs) to get a smooth fit to the profiles. The \siiv\ and \civ\ BELs were fit in \citet{Capellupo11} in order to properly measure the \siiv\ and \civ\ BALs. We do not fit the Ly$\alpha$, \nv, or \ovi\ BALs or adjacent emission lines in the Ly$\alpha$ forest because of the large uncertainties and, specifically, the severe blending of Ly$\alpha$ and \nv\ with each other and possibly with \siiii\ \lam 1206. Also, the \ovi\ BAL is not covered in the Lick 1989.26 spectrum and, therefore, it is not useful for our variability analysis. While we do measure the \pv\ absorption, we do not fit \pv\ emission because it does not have a significant emission line. The black dotted curve in Fig \ref{spectra} shows the pseudo-continuum fit, combining the power-law continuum fit and the fits to the \siiv\ and \civ\ BELs.

The strongest BALs in our data are identified by horizontal solid lines in Fig. \ref{spectra}. Redward of the Ly$\alpha$ emission are the well-studied, aforementioned \siiv\ and \civ\ BALs. Blueward of Ly$\alpha$ are \ovi\ \lam1037, \siv\ \lam\lam1063,1073, \pv, and \nv\ \lam\lam1239,1243 \citep{Leighly09,Baskin13}.

To compare the spectra from different epochs, we use a simple multiplicative shift, with no change to the slope, to match the spectra along regions free of emission and absorption. Fig. \ref{spectra} shows the Lick spectrum from 1989.26 (black curve), with the HST FOS (1994.98; purple curve), HST STIS (2000.29; green curve), SDSS (2006.31; red curve), and KPNO (2010.11; blue curve) spectra overplotted. The \civ\ BAL has a longer wing on the blue side of the profile in the Lick spectrum, and the variability in the BAL extends to higher outflow velocities than in \siiv. There also appears to be variability in \nv\ at $\sim$1140$-$1160\AA, corresponding to the variability in the wing of the \civ\ line. We do not have coverage across \ovi\ in the Lick spectrum, so we cannot compare its variability properties to those of the other lines. We also note that there is variability between the Lick spectrum and the later spectra at $\sim$1580\AA, which may be caused by variability in \feii\ emission lines and/or absorption in these features \citep{Vestergaard05}.

To compare the variability in the different lines, we mark the region of variability in the blue side of the \siiv\ trough with a shaded bar. We then mark the corresponding velocities in the \pv, \nv, and \civ\ BALs with shaded bars, with the widths of the bars adjusted for the different doublet separations in the lines. \civ\ has the narrowest doublet separation at 498 \kms, and \siiv\ and \pv\ have wider doublet separations of 1930 and 2670 \kms, respectively. Fig. \ref{stack} shows the \civ, \siiv, and \pv\ profiles for three of the epochs, namely, Lick 1989.26, SDSS 2006.31, and KPNO 2010.11, shown here again with black, red, and blue curves, respectively. We plot here the normalized spectra, after dividing by the pseudo-continuum fit shown in Fig. \ref{spectra}. The velocity scales in the three panels are defined based on the wavelength of the short-wavelength doublet member for each ion.

Fig. \ref{stack} shows clearly how the profiles of these three lines match in velocity space. The main difference, as mentioned above, is that \civ\ has a more extended wing on the blue side of the profile and the variability extends to higher outflow velocities than in \siiv\ and \pv. The velocities where variability occurred in the blue side of the \siiv\ trough and in the \pv\ trough closely match. The \pv\ line also varies in the same sense as \civ\ and \siiv; all of the lines get weaker in the blue side of the troughs between 1989.26 and the later epochs.

The changes in the \pv\ line profile occur specifically between the Lick and the FOS observations, which are separated by 1.6 yr in the rest-frame of the quasar. The \pv\ line did not vary between any of the later observations presented here. However, in \citet{Capellupo13}, we found variability on time-scales as short as 0.25 yr (91 days) in \civ\ between 2009.22 and 2010.11.

\subsection{Line optical depths and total column density}
\label{tau}

Here we focus on the Lick 1989.26 observation to derive the column densities of the outflowing gas.  We first derive the apparent optical depth versus velocity,  $\tau_{a}(v)$, for the \civ, \siiv, and \pv\ BALs.

\begin{figure}
  \centering
  \includegraphics[width=65mm]{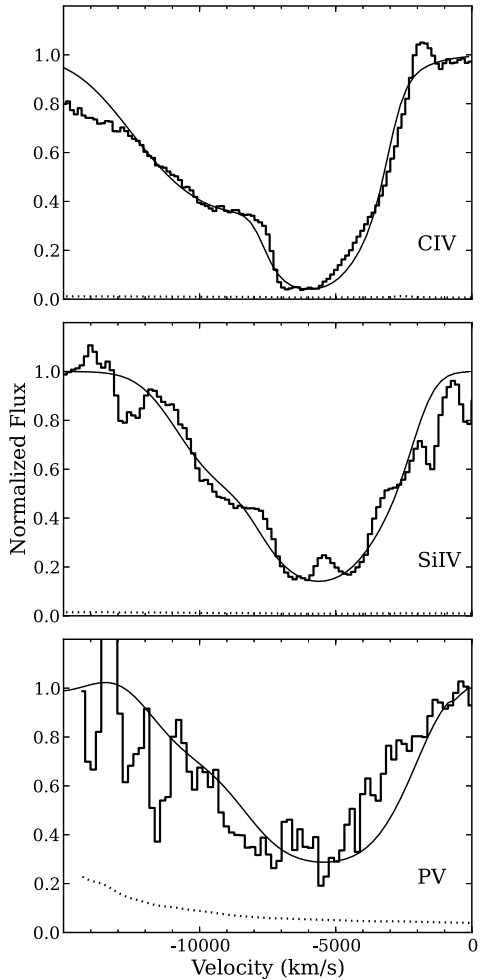}
  \caption{The normalized BAL line profiles for, from top to bottom, \civ, \siiv, and \pv\ for the Lick 1989.26 observation (bold curves). The thin curves show the gaussian fits to the \civ\ and \siiv\ profiles in the top two panels. In the bottom panel, the thin curve is the fit to \pv\ using the optical depth profile for \siiv\ (Fig. \ref{tau_all}).}
  \label{gauss}
\end{figure}

\begin{figure}
  \includegraphics[width=80mm]{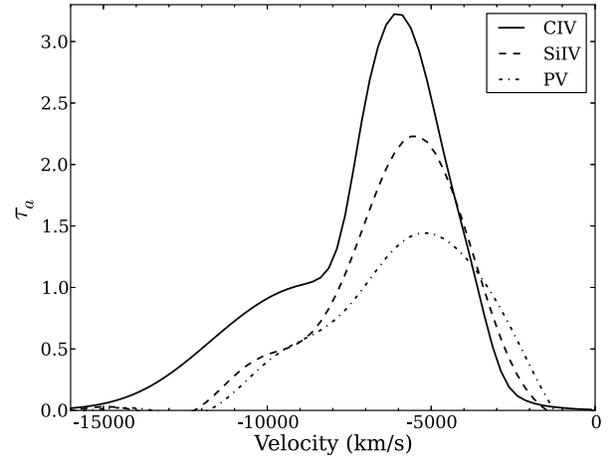}
  \caption{The apparent optical depth profiles for \civ, \siiv, and \pv, after removing the doublet structure in each line. The \civ\ and \siiv\ profiles are derived from the gaussian fits to the residual intensities in the BALs as shown in Fig. \ref{gauss}, and the \pv\ profile is derived based on the \siiv\ optical depth profile.}
  \label{tau_all}
\end{figure}

Using the normalized spectrum, we fit gaussians to the \civ\ and \siiv\ BAL profiles. These gaussians do not represent any physical properties of the gas; they are simply used to define a smooth fit to the profiles (top two panels of Fig. \ref{gauss}). We then use the equation, $I = I_{o}e^{-\tau_{a}}$, to derive $\tau_{a}(v)$ for the \civ\ and \siiv\ lines, where $I_{o} = 1$ for these normalized spectra. The lines we analyze here, \civ, \siiv, and also \pv, are all doublets, so deriving $\tau_{a}(v)$ in this way includes contributions from both transitions at each velocity. We therefore use Equation (2) from \citet{Junkkarinen83} to remove the doublet structure. The solid and dotted curves in Fig. \ref{tau_all} show the run of apparent optical depths versus velocity, $\tau_{a}(v)$, for \civ\ and \siiv, respectively, after removing the doublet structure.

To estimate $\tau_{a}(v)$ for \pv, we first take the optical depth profiles for \civ\ and \siiv\ and broaden the profiles based on the increased doublet separation in \pv. We then calculate $I(v)$ from this adjusted $\tau_{a}(v)$, and fit the resulting \civ\ and \siiv\ intensity profiles to the \pv\ BAL. The \siiv\ $I(v)$ profile provided the best fit to the \pv\ line, with a small adjustment to the amplitude (bottom panel of Fig. \ref{gauss}; see also, \citealt{Junkkarinen97}). We then calculate the \pv\ optical depth profile using this adjusted \siiv\ $I(v)$ profile and $I = I_{o}e^{-\tau_{a}}$, and the result is plotted in Fig. \ref{tau_all} as the dashed-dotted curve. We note that even though the adjusted \siiv\ $I(v)$ profile provides the better fit to the \pv\ BAL, it does overestimate the depth of the absorption on the red side of the trough (Fig. \ref{gauss}). This will have a small effect on the column densities we derive, but it does not significantly affect our final results.

We derive apparent ionic column densities for \civ, \siiv, and \pv\ directly from the $\tau_{a}(v)$ profiles shown in Fig. \ref{tau_all}. These column densities, at least for \civ\ and \siiv, most likely underestimate the true column densities because the lines are probably very saturated (see Section 1 and the Discussion below). We calculate these values for reference using the following equation:
\begin{equation}
    N = \frac{m_{e}c}{\pi e^{2}f\lambda_{0}} \int \tau(v)\ \mathrm{d}v ,
\label{eqn:N}
\end{equation}
where $f$ is the oscillator strength and $\lambda_{0}$ is the laboratory wavelength \citep{Savage91}. We obtain the following values for the ionic column densities: $N_{\mathrm{CIV}} \sim 1.3 \times 10^{16}$ \cmN, $N_{\mathrm{SiIV}} \sim 3.6 \times 10^{15}$ \cmN, and $N_{\mathrm{PV}} \sim 4.0 \times 10^{15}$ \cmN.

Given the small number of lines we measure and the modest constraints they provide on the ionization and column densities, it is sufficient to rely on the published photoionization models by \citet{Hamann98}, \citet{Leighly09,Leighly11}, and \citet{Borguet12} to constrain the true optical depths in the \civ\ BAL and the true total column density, $N_H$, in the Q1413+1143 outflow. The quasar studied by \citet{Borguet12} has much narrower absorption lines than Q1413+1143, particularly in \pv, and the ionic column densities for \civ, \siiv, and \pv\ are roughly an order of magnitude smaller in the quasar they studied. They also have many more lines detected that provide specific constraints on the outflow physical conditions. Based on those constraints, \citet{Borguet12} estimate that the true ratio of \civ\ to \pv\ optical depths should be $\tau_{\mathrm{CIV}}$/$\tau_{\mathrm{PV}}$ $\sim$ 1200 for solar abundances.

\citet{Hamann98}, \citet{Leighly09}, and \citet{Leighly11} consider a wider range of physical conditions in BAL quasars that more closely resemble Q1413+1143, e.g., with broad \pv\ troughs, similar to \civ , and limited ranges of only higher ions detected. In the BAL quasar studied in \citet{Hamann98}, $N_{\mathrm{CIV}}$ is roughly a factor of 2 smaller than in Q1413+1143, and $N_{\mathrm{SiIV}}$ and $N_{\mathrm{PV}}$ are roughly a factor of 4 smaller. \citet{Hamann98} showed that in the limit of low total column densities, where the gas is optically thin throughout the Lyman continuum, the optical depth ratio should be $\tau_{\mathrm{CIV}}$/$\tau_{\mathrm{PV}}$ $\ga$ 800 for solar abundances and any range of ionization conditions. However, those models can be unrealistic if the total columns are too low to yield a detectable \pv\ BAL. \citet{Hamann98} and \citet{Leighly09} estimate that the \pv\ detections in the quasars they studied require minimum total column densities of $N_H\ga 10^{22}$ and $\ga$$10^{22.2}$ cm$^{-2}$, respectively. In that situation, the optical depth ratio should be $\tau_{\mathrm{CIV}}$/$\tau_{\mathrm{PV}}$ $\sim$ 1000 near the column density lower limits and possibly as low as $\sim$100 in extreme circumstances with total columns up to $4\times 10^{23}$ cm$^{-2}$ (e.g., figures 6 in Hamann 1998 and 15 in Leighly et al. 2011). The optical depth ratio in Q1413+1143 should be somewhere in this range. 

To estimate the total column density, we first note that the \pv\ BAL in Q1413+1143 is significantly deeper than the quasars measured by \citet{Leighly09} and especially \citet{Hamann98}, indicating roughly 2--4 times larger \pv\ optical depths and 2--4 times larger \pv\ column densities. If we scale up the column density constraints in \citet{Hamann98} and \citet{Leighly09} by a factor of $\sim$2 based on the stronger \pv\ BAL, we estimate that the total outflow column density in Q1413+1143 is conservatively $\log N_H > 22.3$ cm$^{-2}$.

\section{Discussion}
\label{discuss}

\subsection{Location of the outflow}
\label{loc}

In \citet{Capellupo12} and \citet{Capellupo13} we discussed the two most likely scenarios for explaining the variability in BALs (see also, \citealt{Lundgren07,Wildy14}). The first scenario is a change in the quasar's continuum flux causing global changes in the ionization of the outflowing gas. This could affect the line strengths (optical depths) by changing the column densities in particular ions. There is some evidence for this scenario in outflow lines that vary simultaneously across a wide range in velocities \cite[e.g.,][]{Hamann11,Capellupo12}. However, in Q1413+1143 the detection of a \pv\ BAL indicates that the \civ\ BAL is very saturated. If the optical depth ratio is $\tau_{\mathrm{CIV}}$/$\tau_{\mathrm{PV}}$ $>$ 100 (Section 3.2 above), then the true optical depth across the variable portion of the \civ\ trough is at least $\sim$70 (cf. Figs. 2 and 4). This is much too saturated to be susceptible to changes in the ionization. Similarly, \citet{Hamann08} and \citet{Capellupo13} identify other quasars where a high ratio of \siiv\ to \civ\ absorption strength indicates a saturated variable \civ\ BAL, and they conclude that in these cases, the \civ\ is likely too saturated to be affected by changes in the ionization of the gas.

Instead, the results presented here (and in \citealt{Hamann08} and \citealt{Capellupo13}) strongly support the scenario of clouds moving across our line-of-sight to the quasar continuum source as the cause of the observed variability (see also \citealt{Moe09}, \citealt{Leighly09}, \citealt{Hall11}, \citealt{Vivek12}, and refs. therein). A possible caveat to this interpretation might be that the absorbing region is inhomogeneous, such that it spans a wide range of column densities and line optical depths across our view of the continuum source. In that situation, changes in the continuum flux that change the absorbing region ionization could also change the projected area that has optical depth $\ga$1 in a given line (i.e., the observed covering fraction), without requiring tangential/crossing motions of the gas (see \citealt{Hamann11} and \citealt{Hamann12} for more discussion). However, that possibility seems to be contradicted here by our observations of 1) variability in only portions of the BAL troughs \citep[which is a common feature of BALs][]{Capellupo12,FilizAk13}, and 2) similar variabilities and observed covering fractions in all three lines of \pv , \siiv , and \civ\ that span a factor of $>$100 in optical depth.

We therefore adopt the moving cloud model and use the method described in \citet{Capellupo13} to estimate the location of the gas. This method adopts a simple scenario where a single outflow component of constant ionization and column density crosses our line of sight to the continuum source. The crossing speed we derive from the data depends on the geometry. \citet{Capellupo13} describes two models for the movement of outflow components across the continuum source (see their figure 14 for an illustration). First, a circular disc crosses a larger, circular continuum source along a path that continues through the centre of the background continuum source (the `crossing disc' model). In the second model, the absorber, which has a straight edge, moves across a square continuum source (the `knife edge' model). The `knife edge' model is less realistic, especially given the tendency for variability to occur in small portions of BAL troughs \citep{Gibson08,Capellupo11,FilizAk13}. In the `knife edge' model, the absorber continues to cover a greater percentage of the source over time, until it completely covers the source. This would eventually result in complete absorption at zero intensity. The `crossing disk' model is more realistic, especially in a scenario where separate outflow components, at different velocities, are crossing the background source, causing variability in different portions of the observed spectrum.

We therefore proceed with the `crossing disk' model to calculate the crossing speed of the outflow component. This speed is given by $D_{1500}\sqrt{\Delta A}$/$\Delta$$t$, where $D_{1500}$ is a characteristic diameter for the continuum source at 1500 \AA\ and $\Delta$$A$ is the change in the absorber line-of-sight covering fraction on a time-scale of $\Delta$$t$ \citep{Capellupo13}. For saturated lines like the \civ\ BAL here, $\Delta A$ equals the change in absorption strength relative to a normalized continuum. We calculate a $D_{1500}$ of $\sim$0.004 pc using our estimated bolometric luminosity for Q1413+1143 of $L = 2.4 \times 10^{46}$ ergs s$^{-1}$ (\citealt{Peterson04,Bentz07,Gaskell08}; Hamann \& Simon, in preparation).

For $\Delta$$A$ and $\Delta$$t$, we use results from \citet{Capellupo13} and the current work.
We have more spectra that cover \civ\ than \pv, and the shortest time-scale over which we detect \civ\ variability is 0.25 yr. The portion of the \civ\ BAL that varied had a change in absorption depth of $\Delta$$A$ $\sim$0.09 \citep{Capellupo13}. Since we only observed variability in \pv\ between the Lick 1989.26 and FOS 1994.98 spectra, the maximum variability time-scale for the \pv\ line during the period of our observations is 1.6 yr. The variability in \civ\ on the shorter time-scale of 0.25 yr occurs over a much smaller portion of the BAL than on the longer time-scale of 1.6 yr. Furthermore, in general, variability over 1.6 yr is much more common than on time-scales of 0.25 yr \citep{Capellupo13}. Perhaps, the variability seen on the shorter time-scales is caused by smaller components in the flows at small distances, while the bulk of the flow, represented by the larger, more stable BAL feature, resides further out. We adopt the more conservative approach using 1.6 yr as the $\Delta$$t$ for determining the crossing speed of the outflow, and since the FOS data does not cover the \civ\ absorption line, we adopt the value of $\Delta$$A$ $\sim$0.09. This gives a crossing speed of 750 \kms.

Continuing with the procedure of \citet{Capellupo13}, we assume the crossing speed is rougly equal to the Keplerian rotation speed around an SMBH, with a mass of $M_{BH} \sim 4.8\times10^{8}M_{\sun}$, to get the physical location of the gas. Using the crossing speed based on the `crossing discs' model and the time-scale of 1.6 yr, the distance is $\sim$3.5 pc from the central black hole. Since the variability may have occurred over a shorter time-scale, this distance is likely an upper limit.

\subsection{Energetics of the outflow}
\label{ener}

The energetics of quasar outflows, and thus their potential to affect the host galaxy and large-scale environments via feedback, is still poorly understood. To estimate the energetics, we first need to know the mass of the flow. If we estimate the flow geometry as part of a thin spherical shell, then the total mass is given by
\begin{equation}
    M\ \approx\ 4 100 \, \left(\frac{Q}{15\%}\right) \left(\frac{N_{H}}{2 \times 10^{22}\,\mathrm{cm^{-2}}}
            \right) \left(\frac{R}{3.5\,\mathrm{pc}}\right)^2 \ \ \mathrm{M_{\sun}},
\label{eqn:M}
\end{equation}
where $Q$ is the global covering fraction of the outflow, i.e. the fraction of 4$\pi$ steradians that the flow covers from the point-of-view of the central continuum source, $N_H$ is the total column density of the flow, and $R$ is the distance of the flow from the central SMBH \citep{Hamann00}. We adopt a value of $Q \sim$ 15\% based on the fraction of quasars that exhibit BALs in their spectra (\citealt{Hewett03}; \citealt{Reichard03b}; \citealt{Trump06}; \citealt{Knigge08}; \citealt{Gibson09}).

For most BAL quasars, it is difficult to constrain $N_H$ because few lines are measured and/or the broad profiles blend together and hide the doublet structure in well-studied BALs like \civ. In most cases, the data yield only lower limits on the optical depths and the column densities of the outflowing gas. In this regard, the detection of a low-abundance line such as \pv\ is important because it allows us to estimate the true optical depth in the BALs and better constrain the total column density (see Section 1 and references therein). We adopt the value of $N_H \sim 2 \times 10^{22}$ \cmN, based on the apparent optical depth we derive for \pv\ in Q1413+1143 and the photoionization models of \citet{Hamann98} and \citet{Leighly09} (Section 3.2).

The remaining variable is the radial distance, for which we adopt the value $R\sim 3.5$ pc derived above from the \civ\ variations. Note that Eqn. \ref{eqn:M} provides only a crude approximation because the value of $N_H$ is a lower limit (Section 3.2) while the distance of 3.5 pc is likely an upper limit (Section 4.1). Nonetheless, we can see how changes to these parameters would alter the result. A firm upper limit on the total column in the \civ\ BAL region is $N_H\la 10^{24}$ cm$^{-2}$ to avoid large electron scattering optical depths that would extinguish all of the near-UV flux. Outflows that are Compton thick are also much too mass loaded to be driven out by radiation pressure (see Hamann et al., in prep.). A lower limit on the radial distance is more difficult to establish, but at $R$ much smaller than a parsec, it becomes increasingly difficult to maintain the observed moderate degrees of ionization so close to the continuum source (\citealt{Hamann13}).

To estimate the mass-loss rates, $\dot{M}$, we divide the mass in Eqn. \ref{eqn:M} by a characteristic flow time, $t_{flow} \sim R/v$. We adopt a nominal value of 10 000 \kms\ for the velocity of the flow based on the velocity of the variable interval identified in Figs \ref{spectra} \& \ref{stack}. At a distance of 3.5 pc, $t_{flow}$ is $\sim$340 yr, and the mass-loss rate is therefore 12 M$_{\sun}$ yr$^{-1}$. For comparison, we estimate the black hole mass accretion rate with the equation $L = \eta\dot{M}_{\mathrm{acc}}c^2$, where $\eta$, the efficiency, is nominally 0.1 \citep{Peterson97}. For the luminosity, $L$, we use the bolometric luminosity of $\sim2.4 \times 10^{46}$ ergs s$^{-1}$ (Section \ref{data}), which gives a mass accretion rate, $\dot{M}_{\mathrm{acc}}$, of 4.2 M$_{\sun}$ yr$^{-1}$. Hence, the mass-loss rate through the outflow is roughly 3 times greater than the mass accretion rate onto the central black hole.

The kinetic energy of the outflow, as defined by $K = Mv^{2}/2$, is
\begin{equation}
    K \approx\ 4 \times 10^{54} \, \left(\frac{M}{4100 \, \mathrm{M_{\sun}}}\right)
      \left(\frac{v}{10\,000 \, \mathrm{km\,s^{-1}}}\right)^2 \, \mathrm{erg}.
\label{eqn:K}
\end{equation}
Dividing by the characteristic flow time, $t_{flow}$, gives a time-averaged kinetic luminosity of $\langle L_k\rangle \sim 4 \times 10^{44}$ ergs s$^{-1}$. The ratio of this kinetic energy luminosity to the quasar bolometric luminosity is therefore $\langle L_k\rangle/L \sim 0.02$. This is roughly a factor of 3 smaller than the ratio of $\langle L_k\rangle/L \sim 0.05$ that is typically cited as the ratio necessary for an outflow to be important for feedback (\citealt{Scannapieco04}; \citealt{DiMatteo05}; \citealt{Prochaska09}), but it exceeds the smaller ratio of $\langle L_k\rangle/L \sim$ 0.005 estimated theoretically by \citet{Hopkins10}. As mentioned above, our estimate of the column density is a lower limit, so if the column density is at least a factor of 2.5 greater, at the maximum distance of 3.5 pc, then the ratio of $\langle L_k\rangle/L$ would also exceed the larger threshold of 0.05. Therefore, the outflow might be important for feedback to the host galaxy.

Finally, we point out that the radial distances and kinetic energies derived here, and in other studies of outflow variability \citep{Misawa07,Capellupo11,Capellupo13,Hall11,Hamann11,Hamann13,RodriguezH11,RodriguezH13,Vivek12b}, are smaller than recent results that use excited-state lines to constrain the outflow densities and radial distances (via photoionization models, e.g., \citealt{Moe09,Dunn10, Borguet13, Arav13}). The excited-state absorber studies tend to find distances $\ga$1 kpc and kinetic energy luminosities that are important for feedback (sometimes by a large margin). It is not obvious how this apparent discrepancy between the distance results from variability and excited-state lines can be reconciled. It is possible that BAL outflows exist across a wide range of physical scales, perhaps manifesting themselves differently in different quasars. \citet{Moe09} argue that a variable component of one BAL outflow resides at distances $\la$0.1 pc while another component {\it in the same quasar} with excited-state lines is at $R\sim 3.3$ kpc. It will be an important test of all of these studies to look for variability in BALs known to have excited-state lines and, conversely, to search for excited-state lines to provide density constraints in BALs systems that vary.

\section{Conclusions}

Using the first reported detection of a variable \pv\ BAL, we constrain the mass and energetics of the outflow in Q1413+1143. The detection of \pv\ absorption, which indicates a very saturated \civ\ BAL, and the similar variabilities in the \pv\ and \civ\ BALs imply that the detected variability is caused by the outflow moving across our line-of-sight. This crossing clouds scenario provides a constraint on the distance of the outflow from the central SMBH, and the detection of the \pv\ BAL, along with photoionization models, constrains the true column density of the flow. Together, these quantities allow us to derive an estimate of the mass and energetics of the flow. Our upper limit for the distance of the flow and the lower limit on the column density gives $\langle L_k\rangle/L$ $\sim$ 0.02. This is between the ratio of 0.005 cited by \citet{Hopkins10} and 0.05 determined by earlier work \citep{Scannapieco04,DiMatteo05,Prochaska09} for an outflow to be sufficiently powerful to affect the host galaxy evolution. Therefore, the outflow in Q1413+1143 might be an important feedback mechanism to its host galaxy.

Using large samples of quasars from the ongoing SDSS-III BOSS survey, which includes many quasars at a large enough redshift to cover the \pv\ region, we can search for more detections of \pv\ absorption. We can also look for more cases of variable \pv\ absorption via repeat observations within SDSS-III and follow-up observations with other telescopes. Estimating column densities and distances for flows in more quasars will better able us to assess the typical strengths of these outflows and their viability as a feedback mechanism.

\section*{Acknowledgments}

We thank the referee for helpful comments on the manuscript.
FH acknowledges support from the USA National Science Foundation grant AST-1009628.

\bibliographystyle{mn2e}

\bibliography{full_bibliography}

\bsp

\label{lastpage}

\end{document}